\documentclass[review,3p,times]{elsarticle}%
\usepackage{color}
\usepackage{graphicx}
\usepackage{graphics}
\usepackage{amsmath}%
\usepackage{amsfonts}%
\usepackage{amssymb}%
\usepackage{graphicx}
\usepackage[english]{babel}  %for hyphenation patterns
\usepackage{psfrag} %for graphics, does not work in dvi
\usepackage[T1]{fontenc}
\usepackage[scr=boondoxo,cal=euler]{mathalfa}% for special alphabets; by lazyeqn.sty
\usepackage{tikz}
\usepackage[unicode]{hyperref}
\usepackage{subfig}
\newcommand{\cred}{}

\newcommand{\crev}{}
\newcommand{\bydef}{\,\raise.050ex\hbox{\rm:}\kern-.025em\hbox{\rm=}\,}
\newcommand{\defby}{=\raise.075ex\hbox{\kern-.325em\hbox{\rm:}}\,}
\def\qed{\relax\ifmmode\hskip2em \Box\else\unskip\nobreak\hskip1em $\Box$\fi}
\newcommand {\eps} {\varepsilon} % espilon
\def\eps{\varepsilon}
\begin{document}
\begin{frontmatter}
\title{Tuning frequency band gaps of tensegrity metamaterials with local and global prestress}
%\title{Experimental and numerical study of wave dynamics in tensegrity columns}
%% use optional labels to link authors explicitly to addresses:
%% \author[label1,label2]{<author name>}
%% \address[label1]{<address>}
%% \address[label2]{<address>}

\author[a]{Ada Amendola\corref{mycorrespondingauthor}}
\cortext[mycorrespondingauthor]{Corresponding author}
\address[a]{Department of Civil Engineering, University of Salerno, Via Giovanni Paolo II, 132, 84084 Fisciano (SA), Italy
\\
adaamendola1@unisa.it,f.fraternali@unisa.it},
\author[b]{Anastasiia Krushynska}
\address[b]{
	Department of Physics, University of Turin, Via P. Guria, 1, 10125 Turin, Italy\\
	akrushynska@gmail.com},
\author[c]{Chiara Daraio}
\address[c]{ Engineering and Applied Science, California Institute of Technology, Pasadena, CA 91125, USA\\
	daraio@caltech.edu}
\author[d,e,f]{Nicola M. Pugno}
\address[d]{
	Laboratory of Bio-Inspired and Graphene Nanomechanics, Department of Civil, Environmental and Mechanical Engineering\\ University of Trento, Via Mesiano, 77,
	38123 Trento, Italy\\
	nicola.pugno@unitn.it}
\address[e]{School of Engineering and Materials Science, Queen Mary University of London\\ Mile End Road, London E1 4NS, UK}
\address[f]{Ket Labs, Edoardo Amaldi Foundation, Italian Space Agency\\Via del Politecnico snc, Rome 00133, Italy}	
\author[a]{Fernando Fraternali}
%\address[e]{
%Department of Civil Engineering, University of Salerno, Via Giovanni Paolo II, 132, 84084 Fisciano (SA), Italy\\
%f.fraternali@unisa.it}

%%%%%%%%%%%%%%%%%%%%%%%%ABSTRACT%%%%%%%%%%%
\begin{abstract}
This work studies the acoustic band structure of tensegrity metamaterials, and the possibility to tune the dispersion relation of such systems by playing with local and global prestress variables. Building on established results of the Bloch-Floquet theory, the paper first investigates the linearized response of chains composed of tensegrity units and lumped masses, which undergo small oscillations around an initial equilibrium state. The stiffness of the units in such a state varies with an internal self-stress induced by prestretching the cables forming the tensegrity units, and the global prestress induced by the application of compression forces to the terminal bases.  The given results show that frequency band gaps of monoatomic and biatomic chains can be effectively altered by the fine tuning of local and global prestress parameters, while keeping material properties unchanged.
Numerical results on the wave dynamics of chains under moderately large displacements confirm the presence of frequency band gaps of the examined systems in the elastically hardening regime.
Novel engineering uses of the examined metamaterials are discussed.
\end{abstract}


\end{frontmatter}

%% main text
%%%%%%%%%%%%%%%%%%%%%%%%[SECTION: INTRODUCTION]%%%%%%%%%%%
\section{Introduction}\label{intro}

The research area of linear and weakly nonlinear wave dynamics has devoted much attention to so-called `phononic band gap' theory, which extends the previously investigated theory of photonic band gaps \cite{43,44,Daraio,Phani}. A number of studies have shown that composite materials that feature periodic variations in density and/or wave velocity can display band gaps where the propagation of mechanical waves is forbidden (refer, e.g., to \cite{Phani} and references therein). 
{\cred{Lattice}} metamaterials formed by tensegrity units and lumped masses, which exhibit a peculiar, nonlinear mechanical response are particularly interesting for applications. Such systems are easily tunable{\cred{: either by}} initial self-stress of the units (also referred to as `local' or `internal' prestress), or by changing the precompression of the whole structure (`global' or `external' prestress, refer to  \cite{Skelton_2010}- \cite{Rimoli2017}, and references therein, for an extensive overview).

The research conducted so far in the area of tensegrity metamaterials has revealed that elastically hardening systems support compressive solitary waves and the unusual reflection of waves on material interfaces \cite{JMPS2012,Davini2016}. At the contrary, elastically softening systems support the propagation of rarefaction solitary waves under initially compressive impact loading \cite{JPS20,Experimental,APL2014}. Solitary wave dynamics has been suggested for the construction of a variety of novel acoustic devices, like impact mitigation systems and tunable acoustic lenses. {\cred{Effective impact}} mitigation systems based on tensegrity metamaterials with softening-type response are able to transform compressive disturbances into solitary rarefaction waves with progressively vanishing oscillatory tail  \cite{JPS20,Herbold2012}. Tunable acoustic lenses based on elements with an stiffening response can {\cred{spatially focus}} compression solitary waves {\cred{in different regions of space}} \cite{JMPS2012, Spadoni2010}.

This work investigates translational waves in 1D periodic arrays of tensegrity lattice units alternating with lumped masses, which are shown to be able to control linear elastic waves with arbitrary tunable performance starting from (theoretically) zero frequency. The tuning mechanism relies on variability of an effective stiffness of the tensegrity units by means of applied local and global prestress \cite{JMPS2012}-\cite{APL2014}.
We show that such systems support phononic band gaps, which can be tuned to selected frequency ranges by varying the applied prestress, while keeping material properties of the unit cells unchanged. As compared to granular systems (refer, e.g., to \cite{Nesterenko, Theocharis} and references therein), the internal prestress adds a significant extra feature of tensegrity metamaterials, which can be finely tuned in order to essentially vary the system band gaps.

The structure of the paper is as follows: In Sect. \ref{th1} we model the tensegrity metamaterial as a sequence of masses connected by non-linear springs. We first focus on the linearized mechanical response of a 1D monoatomic lattice (Sect. \ref{th1a}) and then we pass to the analysis of a spring-mass lattice which features springs with two different constants $k_1$ and $k_2$, as a consequence of different states of local and global prestress (Sect. \ref{th1b}).
We show that the dispersion relations of such systems are strongly influenced by the applied levels of prestress.
Numerical results obtained in the geometrically nonlinear regime, which accounts for the actual force-displacement response of the tensegrity units under large or moderately large displacements, confirm the presence of frequency band gaps in the dispersion relation of the analyzed systems  (Sect. \ref{nonlin}). The key mechanical features of the such structures are summarized in Sect. \ref{concl}, where we also suggest future research lines for the design and testing of novel band gap metamaterials with tensegrity architecture.

%%%%%%%%%%%%%%%%%%%%%%%%[SECTION: Dispersion relation of 1D tensegrity chains]%%%%%%%%%%%
\section{Dispersion relation of 1D tensegrity chains}\label{th1}
\begin{figure}[htbp]
	\centering%
	\includegraphics[width=370pt]{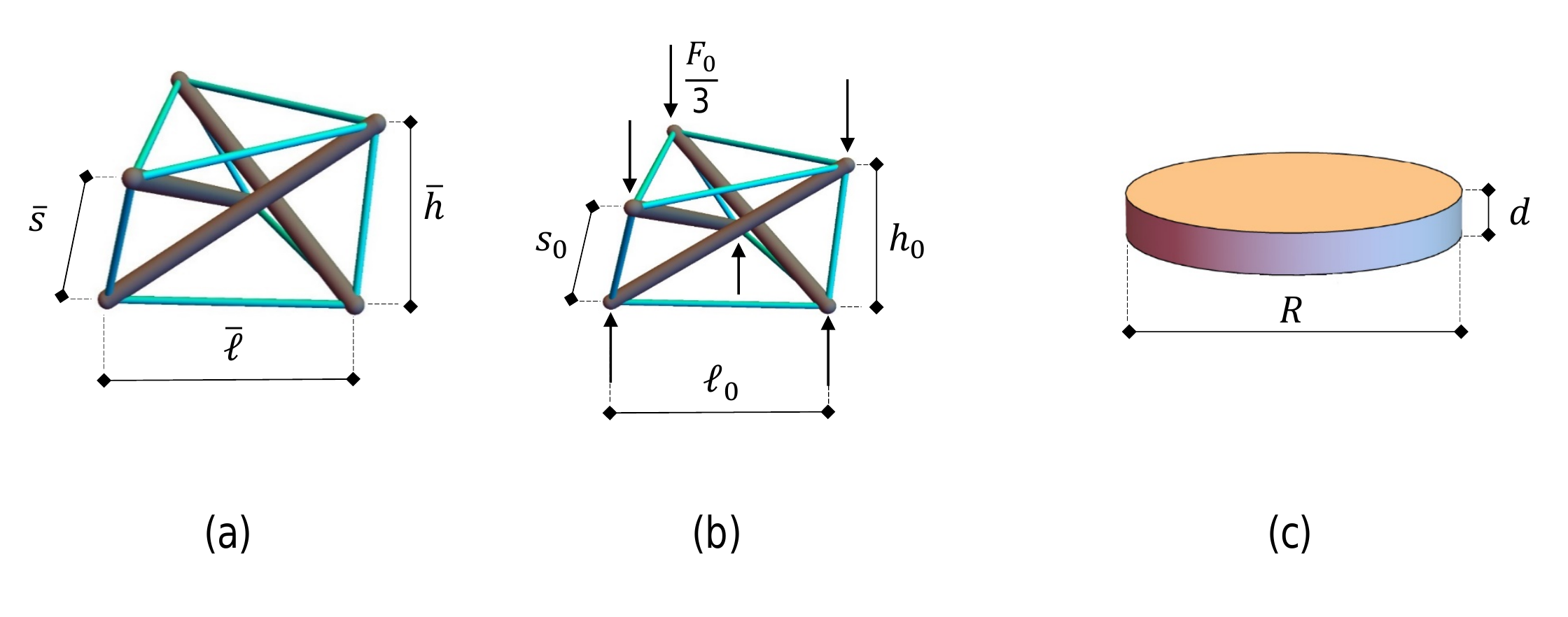}
	%\subfloat[]{\includegraphics[width=110pt]{figures/disco}}
	\caption{Reference configuration of a minimal regular tensegrity prism (a), initial configuration (b) and lumped mass (c).}
	\label{fig:unit}
\end{figure}

The present section studies the dispersion relation of chains obtained by alternating tensegrity units, playing the role of elastic springs, and massive discs, acting as lumped masses. The generic tensegrity unit is composed of the minimal regular tensegrity prism illustrated in Fig. \ref{fig:unit}, which shows two triangular bases composed of members carrying tensile forces (cables or strings), three cross members carrying compressive loads (bars), and three cross-strings. All the strings consist of Spectra fibers, while the bars are made of the titanium alloy Ti6Al4V \cite{Amendola_2015}.

\begin{table}[h]
	\begin{center}
		\begin{tabular}{*{7}{c}}
			%			\hline
			%			Title 1 & Title 2  \\
			\hline
			& $s_N$ & $l_N$  & $R$ & $d$ & $E_b$ & $E_s$\\
			&(mm) & (mm) & (mm) & (mm)& (MPa) &  (MPa) \\
\hline
			& $6.00$ & $8.70$  & $18.66$ & $2.00$ & $120.00$ & $5.48$\\
\hline
\hline
			$\bar{p}$ & 0.00             & 0.01 & 0.02 & 0.05 & 0.09 & 0.10\\
\hline
			$\bar{s}$ \ (mm) &6.000 & 6.060 & 6.120 & 6.300 & 6.540 & 6.600\\
\hline
			$\bar{\ell}$ \ (mm) & 8.700 & 8.773 & 8.845 & 9.061 & 9.346 & 9.417\\
\hline
			$\bar{b}$ \ (mm) & 11.108 & 11.207 & 11.305& 11.597 & 11.985& 12.081\\
\hline
			$\bar{h}$ \ (mm) & 5.407 & 5.463& 5.519 & 5.688 & 5.913 & 5.969\\
		
			\hline
		\end{tabular}
	\end{center}
	\caption{Geometrical and mechanical properties of the Mono 1 and Mono 2 units.}\label{param}
\end{table}
 
The chain is uniformly axially loaded and we assume that the lattice unit cells are in frictionless contact with the lumped masses \cite{APL2014}, so that twisting moments are not transferred to the lumped masses, which therefore move only axially.
The following examines chains equipped with tensegrity prisms featuring identical geometrical properties in the natural (stress-free) configuration and identical mechanical properties, with possibly different mechanical response up to the value of the applied prestress.	
%On assuming that the units and the lumped masses are equal each other, such a system can be described as a 1D monoatomic chain. 

The properties of the considered tensegrity unit are given in Tab. \ref{param}. Hereafter, we use the symbols $s$, $l$, and $b$ to denote the current lenghts of the cross-strings, the base strings and the bars, respectively, and let $h$ denote the height of the unit. In addition, we let $R$ and $d$ indicate the radius and the thickness of the discs interposed between the units, and make use of the symbols $E_b$ and $E_s$ to denote the Young moduli of the bars and the strings, respectively. 
A state of local prestress (or selfstress) is applied to the units in the reference configuration of the chain under zero external force, which shows the two terminal bases of the genric unit rotated against each other at a twisting angle of $5/6$ $\pi$ \cite{JPS20}. Such an internal state of prestress follows from the action of  a self-equilibrated set of axial forces in the prism members, and can be usefully characterized  though the prestrain of the cross-strings $p$ \cite{JPS20}. 
 Additionally, the chain is initially loaded with a static precompression force $F_0$ (total axial force applied to the terminal bases), which induces a state of global prestress to the structure (the initial configuration).
%The initial equilibrium configuration of the chain is characterized by the action of a precompression force $F_0$, which axially deforms the reference configurazion, and produces a state of global prestress of the system.
We agree to denote all the quantities referred to the reference configuration by a superimposed dash, and the quantities referred to the initial configuration by the subscript "0" (cf. Fig. \ref{fig:unit}). {\crev{The total mass $M$ of a unit cell is evaluated as the sum of the disk's mass ($m$) and the prism's mass ($m_0$)}}. We set $m_0= 0.083$ g and $m=300 \ m_0= 24.89$ g, so that the chain can be described as as a system of point masses connected by massless springs ($m>>m_0$ \cite{APL2014}). 

As we shall see later on, the axial force $F$ vs. axial displacement $\delta$ response of the generic unit is markedly nonlinear, due to geometric effects, and exhibits zero slope at the origin under zero prestress \cite{JPS20}. The study presented in the current section linearizes such a response near the initial configuration ($\delta_0, F_0$), by describing the tensegrity unit as a linear spring with stiffness constant $k$ equal to the slope of the tangent line to the $F-\delta$ curve (tangent axial stiffness). The results presented hereafter are therefore valid for (infinitesimally) small oscillations of the system with respect to the initial configuration. 
%We refer the reader to Sect. \ref{nonlin} for a study of the dispersion relation of a tensegrity chain in the nonlinear regime, which is characterized by moderately large oscillations of the system from the initial configuration.

%%%%%%%%%%%%%%%%%%%%%%%%[SECTION: Monoatomic chain]%%%%%%%%%%%
\subsection{Monoatomic chain}\label{th1a}
\begin{figure}[bth] 
\centerline{
\begin{tabular}{c c}
\includegraphics[width=350pt]{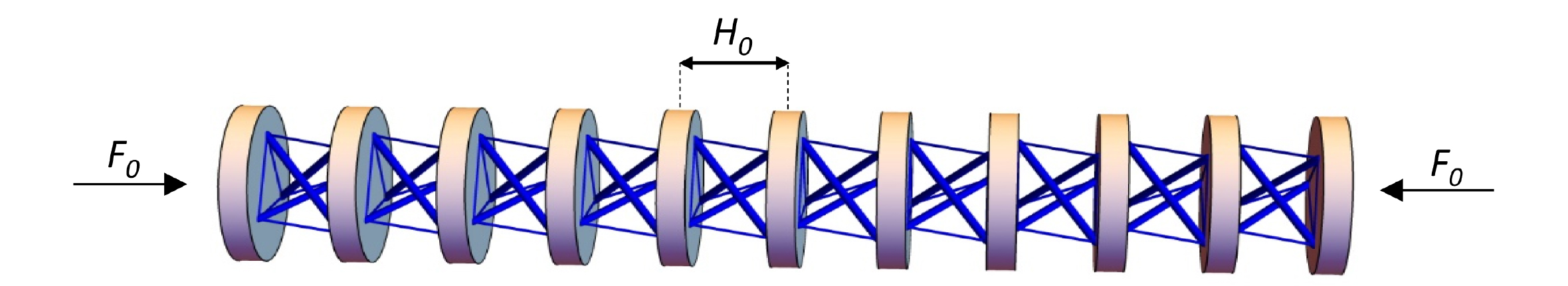}\\
\includegraphics[width=350pt]{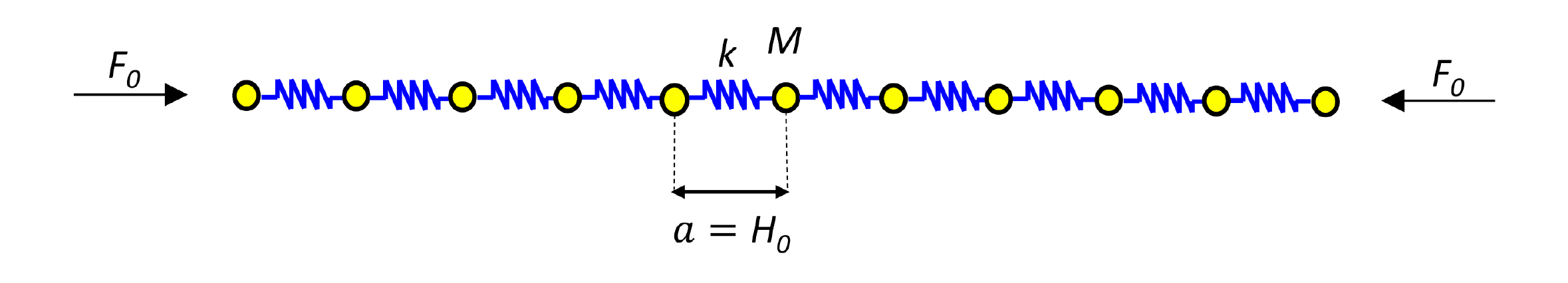}\\
%\multicolumn{2}{c}{\includegraphics[height=5cm]{./Graphics/G_Z14_abcd.pdf}}\\
\end{tabular}
}
\caption{Monoatomic chain: physical model (top) and mass-spring model (bottom).}
\label{fig:monoat}
\end{figure}

Our first goal is to study the band structure of a tensegrity chain by using available results for monoatomic structures \cite{ashmer}. 
%We remark that the mechanical response of the units depends on the nature of the materials employed for bars and strings and on the pre-stress that is applied to the structure in the reference configuration, which shows the two terminal bases rotated against each other at a twisting angle of $5/6$ $\pi$ \cite{27}. 
We analyze a monoatomic tensegrity chain in the form of a sequence of masses connected with linear springs of the same stiffness constant $k$ (Fig. \ref{fig:monoat}). 
We define the distance between the masses as $H_0=h_0+d$, where $h_0$ is the height of the prism in the initial configuration. The quantity $H_0$ corresponds to the unit cell size `$a$'.
%The spring constant $k$ is given by the tangent axial stiffness in the equilibrium configuration, as it will be detailed in the sequel. 

\begin{figure}[htbp]
	\centering%
	\subfloat[]{\includegraphics[width=190pt]{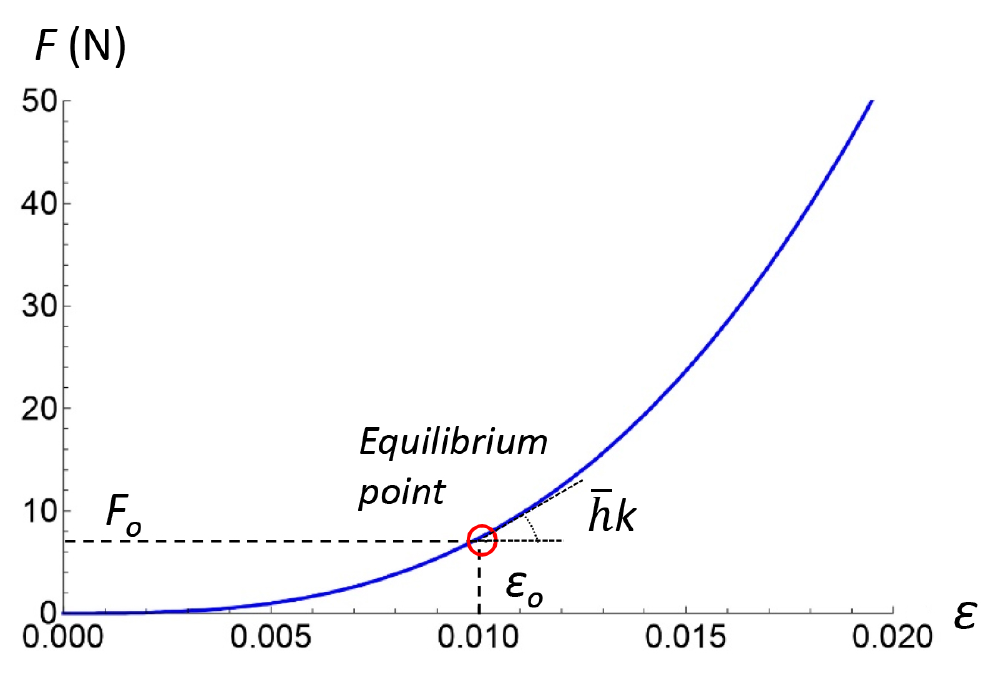}}
	\subfloat[]{\includegraphics[width=190pt]{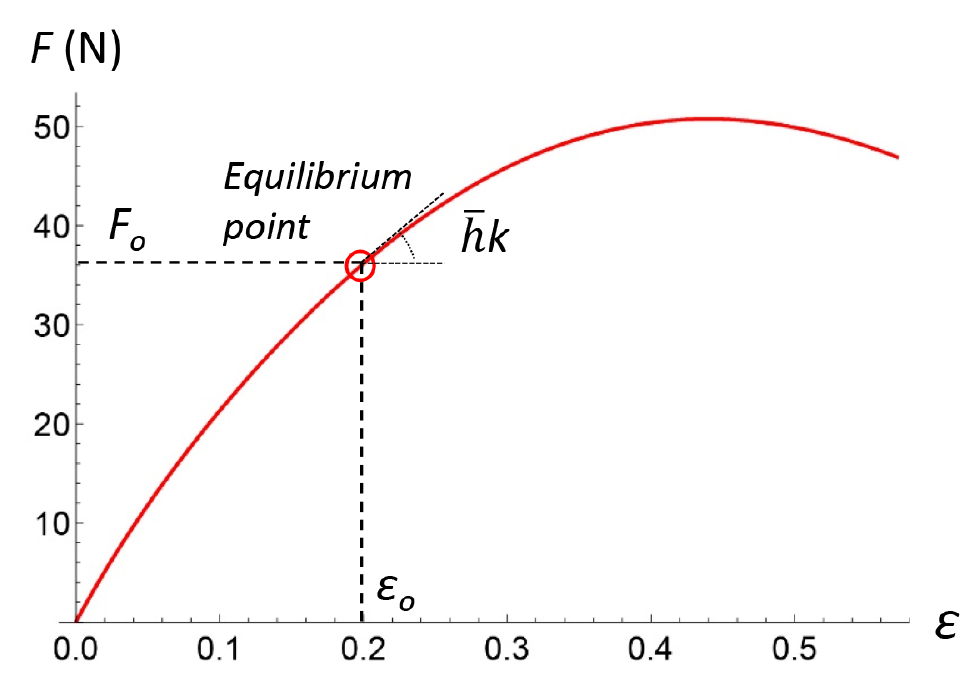}}
	\caption{$F-\eps$ curves of systems Mono 1 (a), and Mono 2 (b).}
	\label{fig:Feps}
\end{figure}

Let $\delta=\bar{h}-h_0$ denote the axial displacement from the reference configuration, and let $\eps = \delta/\bar{h}$ denote the corresponding axial strain (positive when the prism is compressed). 
We examine two different monoatomic chains: `Mono 1' with equilibrium point along a branch of the force-strain curve characterized by stiffening response and relatively low tangent stiffness (Fig. \ref{fig:Feps}a: $F_0=F_{0_1}=7.31$ N, $\eps_{0}=\eps_{0_1}=0.010$, $k=k_1=394.89$ N/m); and `Mono 2' with an equilibrium point along a branch of the force-strain curve characterized by  softening response and relatively high tangent stiffness (Fig. \ref{fig:Feps}b: $F_0=F_{0_2}=36.14$ N, $\eps_{0}=\eps_{0_2}=0.199$, $k=k_2=2165.30$ N/m, cf. Tab. \ref{param}). The nonlinear nature of the $F$ - $\eps$ curves exhibited by the systems under consideration is clearly visible in Fig. \ref{fig:Feps}. As explained above, we linearize such a response near the initial configuration marked by circles in Fig. \ref{fig:Feps}, to examine infinitesimally small oscillations of the chains Mono 1 and Mono 2.
Note that the tangent stiffness $k$ at the initial state is equal to the slope of the local tangent line to the $F-\eps$ curve multiplied by the equilibrium height $\bar{h}$ (Fig. \ref{fig:Feps}).

\begin{table}[h]
	\begin{center}
		\begin{tabular}{*{9}{c}}
			%			\hline
			%			Title 1 & Title 2  \\
			\hline
			& 
			$\bar{p}$   &   
			$\eps_{0}$     & 
			$s_{0}$  & 
			$\ell{_0}$  &  
			$h_{0}$ & 
			$a$  & 
			$\phi_{0}$   & 
			$k$ 
					\\
					& 
			(\%)  &   
			(\%)   & 
			(mm) & 
			 (mm) &  
			(mm) & 
			(mm) & 
			(rad) & 
			(kN/m)
					\\
			% & &  &  \\
\hline
			$\mbox{System}\ 1$ &0&1& 6.001& 8.701&5.353&7.353&2.641&0.39
			%$1*10^{-6}$  & $5.407$ & $8.700$
			\\
\hline
			$\mbox{System}\ 2$ &5&20& 6.159&9.310 &4.551&6.551&2.887&21.65 
			%$5*10^{-2}$  & $5.688$ & $9.061$
			\\
		
			\hline
		\end{tabular}
	\end{center}
	\caption{Geometrical and mechanical properties of the systems shown in Fig. \ref{fig:Feps}.}\label{param1}
\end{table}

\begin{figure}[htbp] 
	\centering
	\includegraphics[width=250pt]{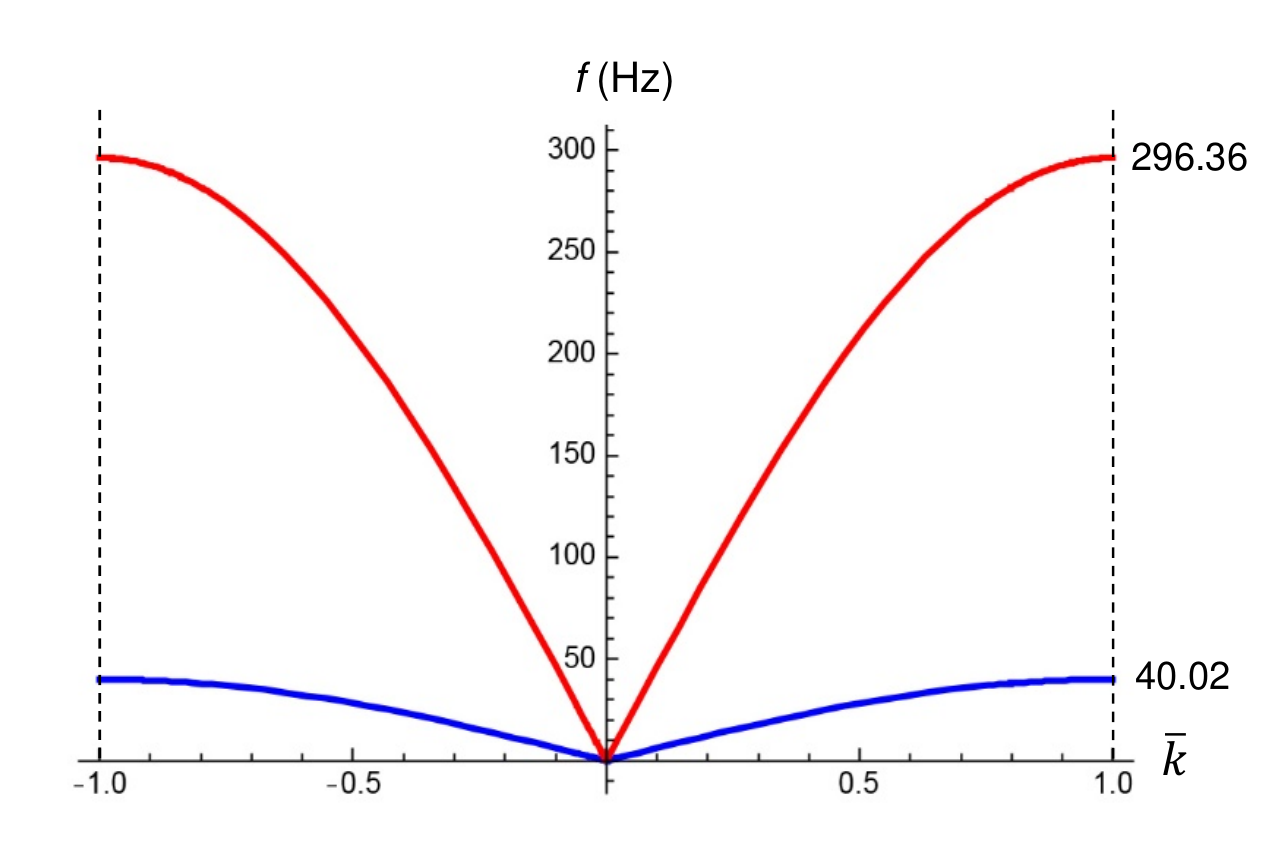}
	\caption{Dispersion relation in the first Brillouin zone for monoatomic chains equipped with prisms corresponding to systems Mono 1 (bottom, blue) and Mono 2 (top, red) in Fig. \ref{fig:Feps}.}
	\label{fig:monodispersion}
\end{figure}

On applying the Bloch-Floquet theory presented in Refs. \cite{ashmer,Herbold} for one-dimensional periodic systems, we obtain a single dispersion curve (the so-called `acoustic branch') for the Mono 1 and Mono 2 systems, as illustrated in Fig. \ref{fig:monodispersion}. Here, $f$ denotes the wave frequency and $\bar{\kappa}$ denotes the normalized wavenumber $\kappa \ \pi / a$.
Let us focus our attention on the high-symmetry points of the first  Brillouin zone ($\bar{\kappa}=\pm 1$), at which the dispersion curves reach the limiting frequencies $f=40.03$ Hz for Mono 1, and $f=296.36$ Hz for Mono 2.
{\cred{The above frequencies (or band edges) mark the upper bound of the transmission region}} \cite{ashmer}. {\crev{We highlight a  $\approx 600 \%$ increase of the lower band gap frequency, when passing from Mono 1 ($\bar{p}=0, \eps_{0}= 1 \%$) to Mono 2  ($\bar{p}=5 \%, \eps_{0}= 20 \%$)}}. Therefore, it is clear that one can markedly change the dynamics of mechanical waves in such systems by playing with the value of the applied, local and/or global, prestress.

%%%%%%%%%%%%%%%%%%%%%%%%[SECTION: biatomic chain]%%%%%%%%%%%
\subsection{Diatomic chain}\label{th1b}

\begin{figure}[bth] 
\centerline{
\begin{tabular}{c c}
\includegraphics[width=350pt]{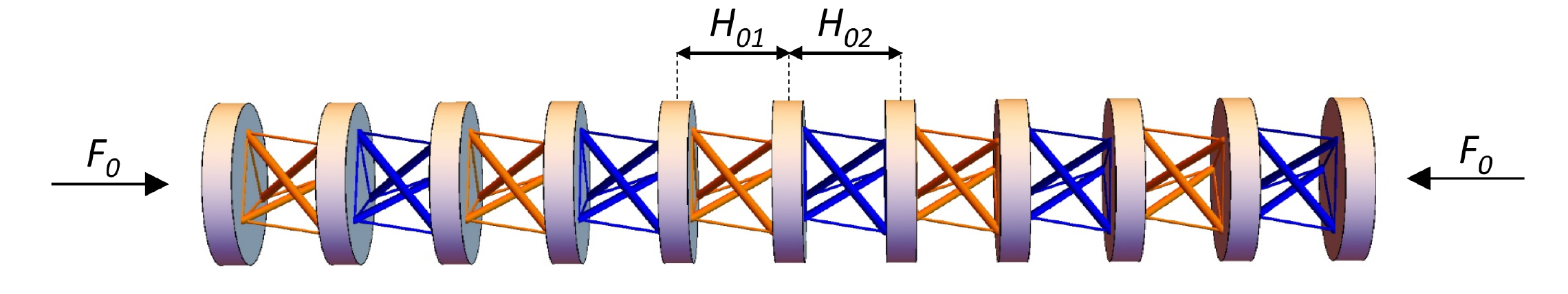}\\
\includegraphics[width=350pt]{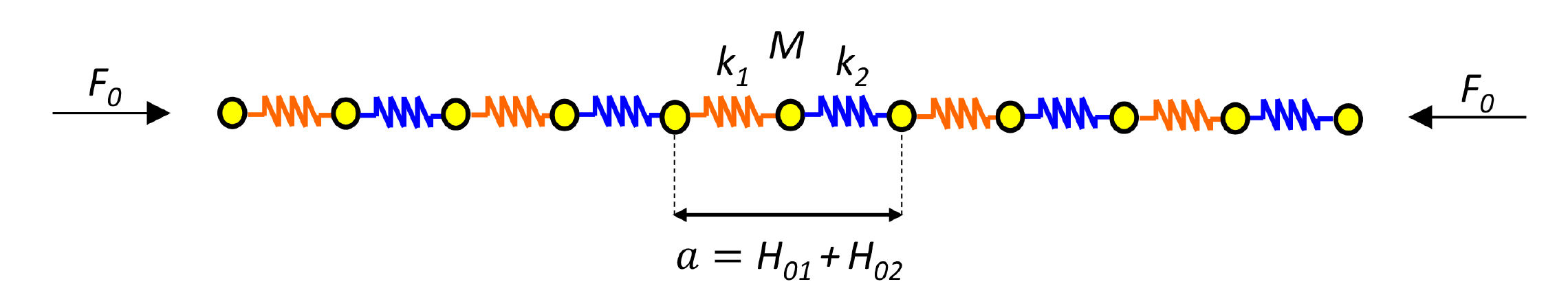}\\
%\multicolumn{2}{c}{\includegraphics[height=5cm]{./Graphics/G_Z14_abcd.pdf}}\\
\end{tabular}
}
\caption{biatomic chain: physical model (top) and mass-spring model (bottom).}
\label{fig:biatomic}
\end{figure}

We now examine the band structure of a biatomic tensegrity chain (Fig. \ref{fig:biatomic}, top), which is modeled by a sequence of identical lumped masses connected by linear springs with alternationg constants $k_1$ and $k_2$, where $k_1< k_2$ (Fig. \ref{fig:biatomic}, bottom). The spring with the constant $k_1$ refers to the 1D model of a tensegrity prism of height `$h_{01}$' at the initial configuration (`soft' spring), while the spring with the constant $k_2$ refers to a prism of height `$h_{02}$'  (`hard' spring).

We define the distance between the two masses connected to the softer prism as $H_{01}=h_{01}+d$. Using the same notation,  $H_{02}$ is equal to $h_{02}+d$. 
Consequently, the unit cell size of the mass-spring model can be defined as $a=H_{01}+H_{02}$ (Fig. \ref{fig:biatomic}).
As in the case of the monoatomic chain, all the prisms forming the biatomic chain are characterized by identical geometric properties in the stress-free configuration, as well as identical material properties, and differ only by the value of the local and global prestress (refer to Tab. \ref{param} for geometric and material properties). 
By fixing the values of the hard spring constant $k_2$ and the applied external precompression force $F_0$, we study the variation of the mechanical properties of the structure with the soft spring constant $k_1$.

\begin{figure}[htbp] 
	\centering
	\includegraphics[width=310pt]{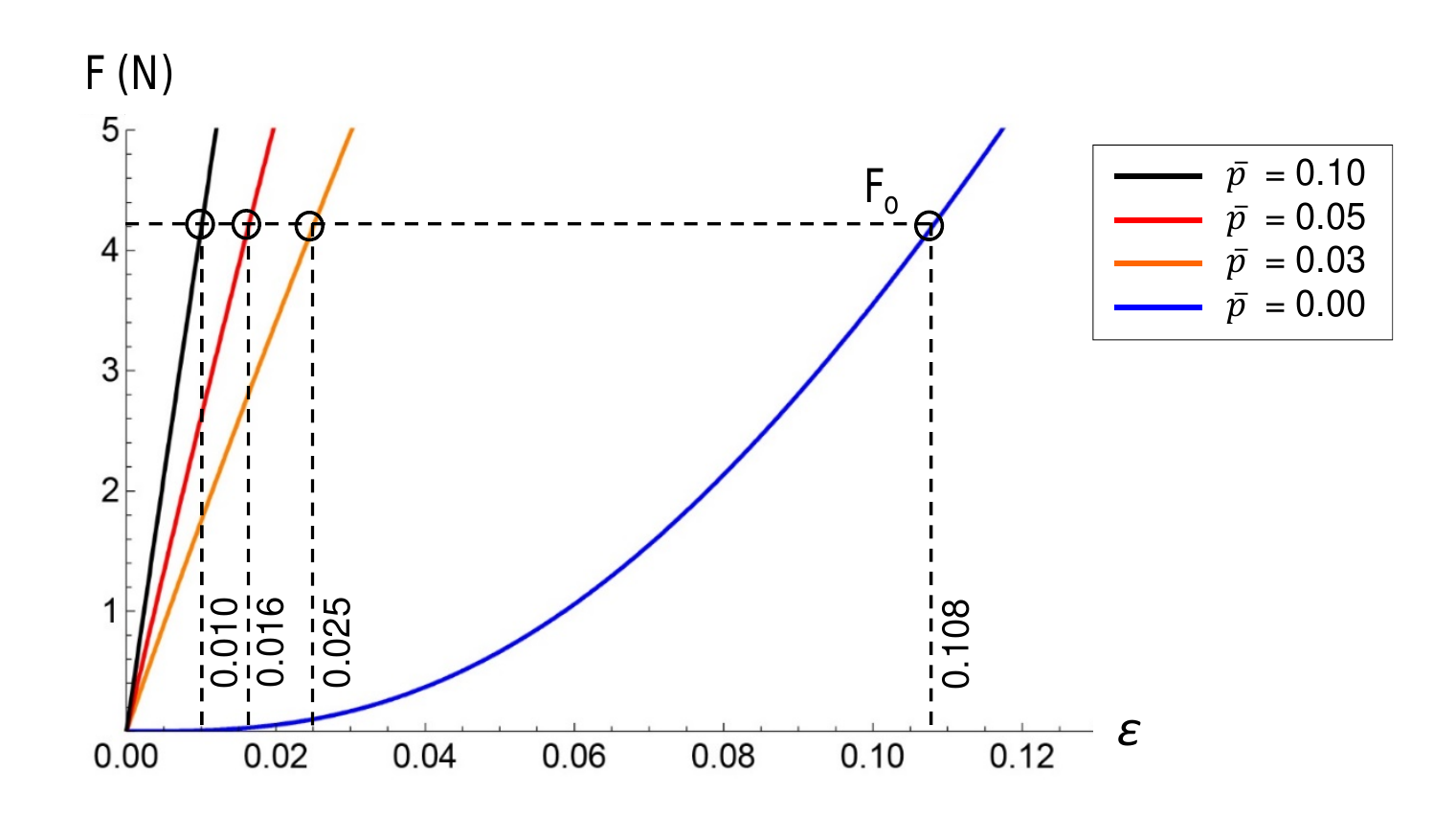}
	\caption{$F-\eps$ curves of various units of biatomic chians for different values of $\bar{p}$ and $F_0=4.21$ N.}
	\label{fig:Fepsbiat}
\end{figure}

We first analyze the case with $F_0=4.21$ N and constant $k_2=68.33$ kN/m. The latter refers to the tangent stiffness at the equilibrium point of a tensegrity prism subject to an internal prestrain $\bar{p}=0.1$ and an external prestrain $\eps_0=0.01$. 
Tab. \ref{biatparameters} provides a complete list of mechanical and geometrical parameters of such a unit, together with the analogous properties of different soft units taken into consideration for various values of $\bar{p}$ and $\eps_0$.
The $F$ - $\eps$ responses of the soft and hard units are illustrated in Fig. \ref{fig:Fepsbiat}. 
%When $F_0$ is kept constant, we notice that higher values of the local and global precompression parameters $\bar{p}$ and $\eps_{0}$ are needed to get softer units.

%inserisci valori giusti tabella
\begin{table}[h]
	\begin{center}
		\begin{tabular}{{c}|{c}{c}{c}|{c}}
		%\hline
%& \multicolumn{4}{c}{\textit{fully elastic model}}  \\
%\hline
& \multicolumn{3}{c|}{\textit{Soft}}  & \multicolumn{1}{c}{\textit{Hard}} \\
\hline
			$\bar{p}$           & 0.000   & 0.030     & 0.050     & 0.100 \\ 
\hline
			$\eps_{0}$         & 0.108  & 0.025     & 0.016   &0.010\\
\hline
			$k_{}$\ (kN/m)   & 15.463 & 28.257   & 43.107 & 68.333\\
\hline
			$s_0$\ (mm)        &6.027    & 6.161    & 6.280   &6.578\\
\hline
			${\ell}_0$ \ (mm) & 8.776   & 8.933   & 9.075    &9.430 \\
\hline
			$b_0$ \ (mm)       & 11.107 & 11.402  & 11.597 &12.081 \\
\hline
			$h_0$ \ (mm)        & 4.823   & 5.436   & 5.595   &5.910\\
\hline
			
		\end{tabular}
	\end{center}
	\caption{Geometrical and mechanical properties of the equilibrium points of soft and hard units of biatomic chains, for $F_0=4.21$ N}. \label{biatparameters}
\end{table}

\begin{figure}[htbp] 
	\centering
	\includegraphics[width=350pt]{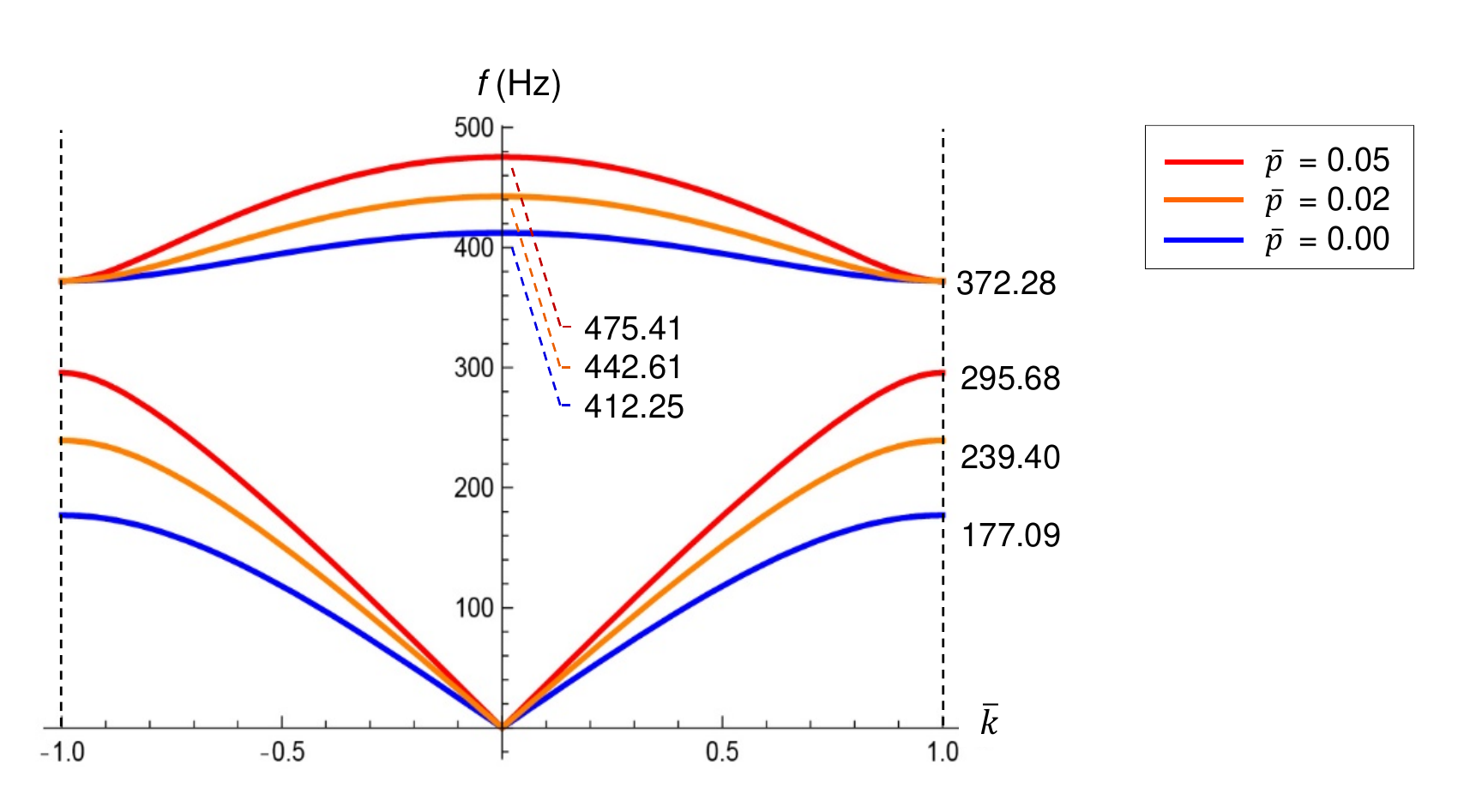}
	\caption{Dispersion relation in the first Brillouin zone for a biatomic chain under $F_0=4.21$ N and different values of $\bar{p}$.}
	\label{fig:biatdispersion}
\end{figure}

Fig. \ref{fig:biatdispersion} shows the two branches of the dispersion relation (acoustic and optical branches, respectively), which were obtained for each analyzed biatomic chain through the theory presented in \cite {ashmer}. Such a dispersion relation reveals the presence of a band gap between  the cut-off frequencies of the acoustic and optical branches, i.e.: $f_1=\sqrt{{2k_1/M}}$ and $f_2=\sqrt{2k_2/M}$, for $\bar{k}=1$  \cite {ashmer}.
By keeping
$\bar{p}=0.100$ and $\eps_{0}=0.010$ in the hard units and varying such prestress parameters in the soft unit, one can get a fixed upper bound at 372.28 Hz and tune the lower band gap bound to 177.09 Hz, 239.40 Hz, and 295.68 Hz when the couple $\{ \bar{p}, \ \eps_{0} \}$ is  respectively equal to $\{ 0.000, \ 0.108 \}$,  $\{ 0.030, \ 0.025 \}$, and  $\{ 0.050, \ 0.016 \}$ in the soft units.
Such results (Fig. \ref{fig:biatdispersion}) confirm that the band gap frequencies in biatomic tensegrity chains can be effectively tuned by playing on the local and global prestress, similarly to the monoatomic systems.

%\begin{figure}[htbp] 
%	\centering
%	\includegraphics[width=260pt]{figures/Fig_8}
%	\caption{$F-\eps$ curves .. $F_0=2.48$ kN.}
%	\label{fig:soft_rig}
%\end{figure}

\begin{figure}[htbp] 
	\centering
	\includegraphics[width=260pt]{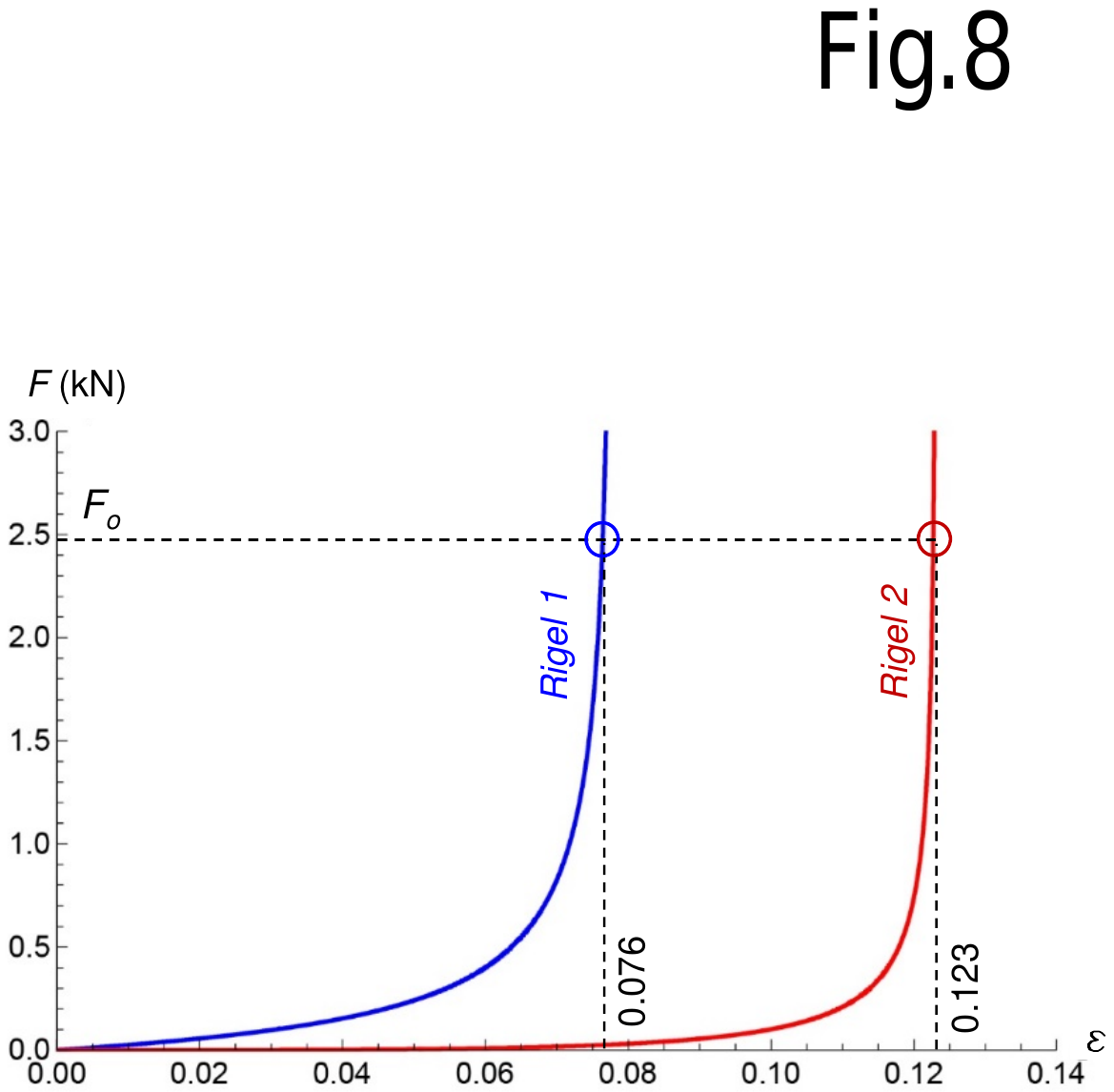}
	\caption{$F-\eps$ curves of the rigid-elastic units under $F_0=2.48$ kN.}
	\label{fig:Fepsbiatrigel}
\end{figure}

The biatomic chains examined in Fig. \ref{fig:biatdispersion} show band gap frequencies between the acoustic and optical branches  within the audible range (20 Hz to 20kHz). This result is a consequence of the particular choice of the prisms and masses forming such systems, and can be generalized to hypersonic band gap system by using, e.g., tensegrity units consisting of prisms equipped with rigid bases and bars (rigid-elastic units \cite{JPS20}). 
Table \ref{biatparametersrigel} shows the geometrical and mechanical properties of two rigid-elastic tensegrity units (named `Rigel 1' and `Rigel 2') that exhibit the same stress-free configuration and identical cross-string material of the fully elastic prisms illustrated in Tab. \ref{param1}. The  axial force vs. axial strain responses of such units illustrated in \ref{fig:Fepsbiatrigel} highlights a locking-type response in correspondence to the limiting configuration with $\phi = \pi$, when the bars touch each other \cite{JPS20}.
The dispersion relation of a biatomic chain equipped with the rigid-elastic units is shown in Fig. \ref{fig:Fepsbiatrigel}. It is seen that no waves can propagate along the chain within the first band gap region 16.84-33.12 kHz, which extends above the audible frequency range.

\begin{table}[h]
	\begin{center}
		\begin{tabular}{{c}|{c}|{c}}

& \textit{Rigel 1}  & \textit{Rigel 2} \\
\hline
			$\bar{p}$           & 0.200   & 0.000      \\ 
\hline
			$\eps_{0}$         & 0.076  & 0.125   \\
\hline
			$k_{}$\ (MN/m)   & $139.866$ & $540.754$  \\
\hline
			$s_0$\ (mm)        &7.782    & 6.810   \\
\hline
			${\ell}_0$ \ (mm)  & 8.700   & 8.700    \\
\hline
			$b_0$ \ (mm)        & 11.800 & 11.108 \\
\hline
			$h_0$ \ (mm)        & 6.201   & 4.744   \\
\hline
			
		\end{tabular}
	\end{center}
	\caption{Geometrical and mechanical properties of the equilibrium points of units equipped with rigid bases under $F_0=2.48$ kN}. \label{biatparametersrigel}
\end{table}

\begin{figure}[htbp] 
	\centering
	\includegraphics[width=260pt]{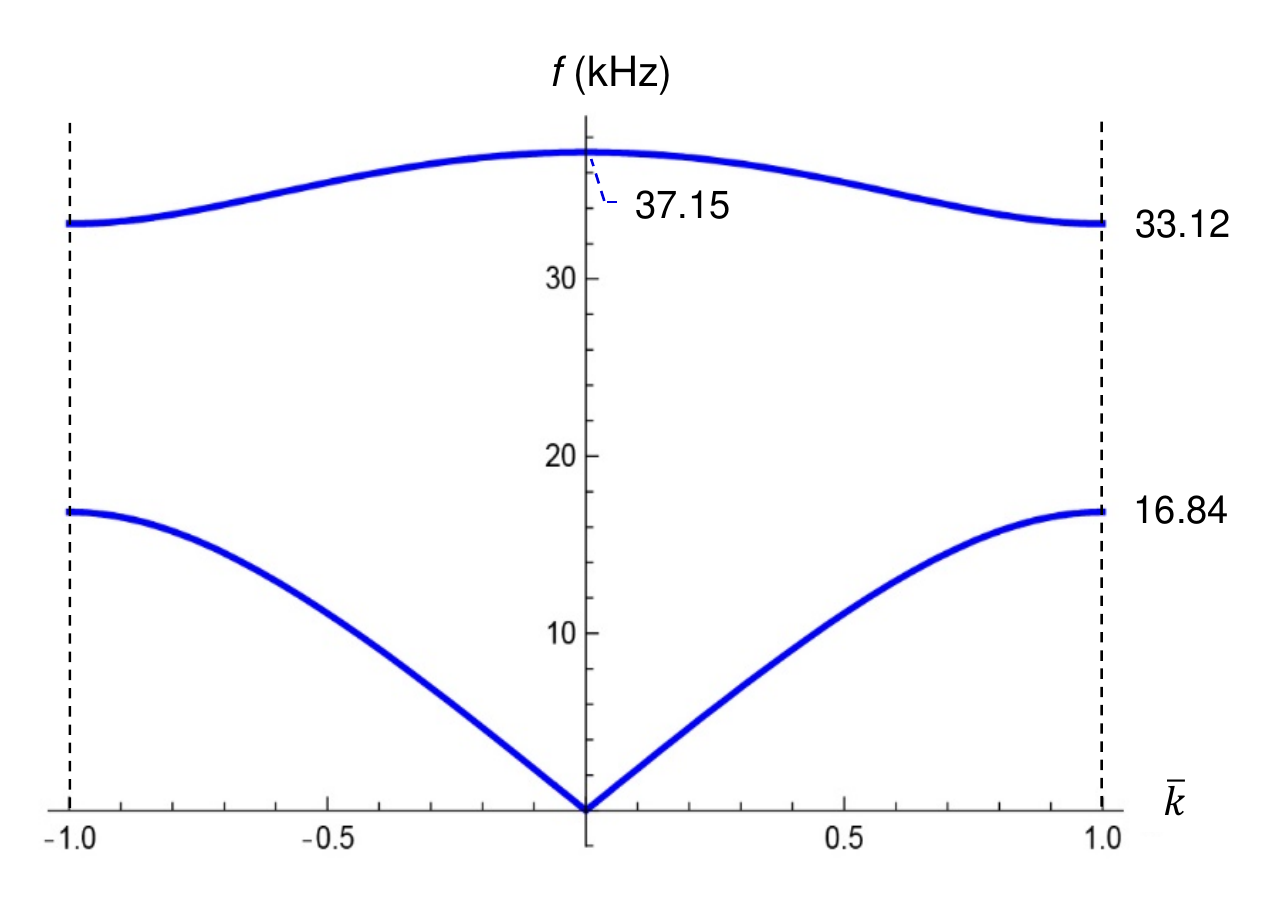}
	\caption{Dispersion relation in the first Brillouin zone for a biatomic chain equipped with Rigel 1 and Rigel 2 units under $F_0=2.48$ kN.}
	\label{fig:biatdispersionrigel}
\end{figure}

%By applying a global precompression force that sets the equilibrium point of such a unit close  to the locking regime \cite{27}, one can maximize the band gap region by letting $k_2$ tend to infinity, which implies $f_2 \rightarrow \infty$, assuming  $M$ constant. 
%With reference to the prisms examined in the present studies, one can, e.g., set....

%%%%%%%%%%%%%%%%%%%%%%%%[SECTION: Dispersion relation of 3D tensegrity columns]%%%
\section{Frequency band gaps under moderately large incremental strains}\label{nonlin}

We have already observed that the actual $F$ vs. $\eps$ curves examined in Sect. \ref{th1} are markedly nonlinear, due to geometric (i.e., large displacement) effects
\cite{JPS20}-\cite{APL2014}  (cf.  Fig. \ref{fig:Feps}). 
%As anticipated, we hereafter linearize such a response near the initial configurations marked by circles in Fig. \ref{fig:Feps}, on examining small oscillations of Systems 1 and 2 from such equilibrium states.
%It is worth noting that the tangent stiffness $k$ at the initial state is equal to the slope of the local tangent line to the $F-\eps$ curve multiplied by the equilibrium height $\bar{h}$ (Fig. \ref{fig:Feps}).
We now want to numerically study the phenomenon of wave attenuation in the Mono 1 chain analyzed in Sect. \ref{th1}, by accounting for the nonlinear response under moderately large incremental strains from the equilibrium configuration.
Such a  study is conducted by perturbing the equilibrium configuration by sinusoidal time-displacement input applied to the first unit of a chain composed of 100 masses. The amplitude of the applied displacement input is set equal to 0.03 mm,  which gives raise to an incremental strain $|\Delta \eps|_{} \approx 0.6 \%$ from the initial equilibrium point (cf. Tab. \ref{param1}), and nearly reduces to zero the static precompression force $F_0=7.31$ N, when applied in tension (cf. Fig. \ref{fig:Feps}(a)). The excitation frequency $f_{in}$ ranges between 20 Hz and 80 Hz, and thus, includes the bandgap edge $40.03$ Hz of the  Mono 1 chain in the linear regime, which is hereafter referred to as `linear Mono 1' chain (Fig. \ref{fig:monodispersion}).
The present study accounts for the nonlinear $F$ vs. $\eps$ law depicted in Fig. \ref{fig:Feps}(a), and is conducted through the particle dynamics code described in Refs. \cite{JMPS2012,54,38}. The given numerical results assume a time integration step equal to $10^{-3} /{f_{in}}$, which is significantly lower than the oscillation period of the linearized unit ($T_0 \approx 0.003$ s, cf. Ref. \cite{APL2014}).
% in all the examined cases. Note that all the traveling pulses hereafter examined exhibit characteristic times $T_w$ much greater than $T_0$ \cite{APL2014}.
%It is worth noting that the applied input induces moderately large incremental strains accompanied by stiffening-type response (cf. Fig. \ref{fig:Feps}(a)).

\begin{figure}[htbp] 
	\centering
	\includegraphics[width=440pt]{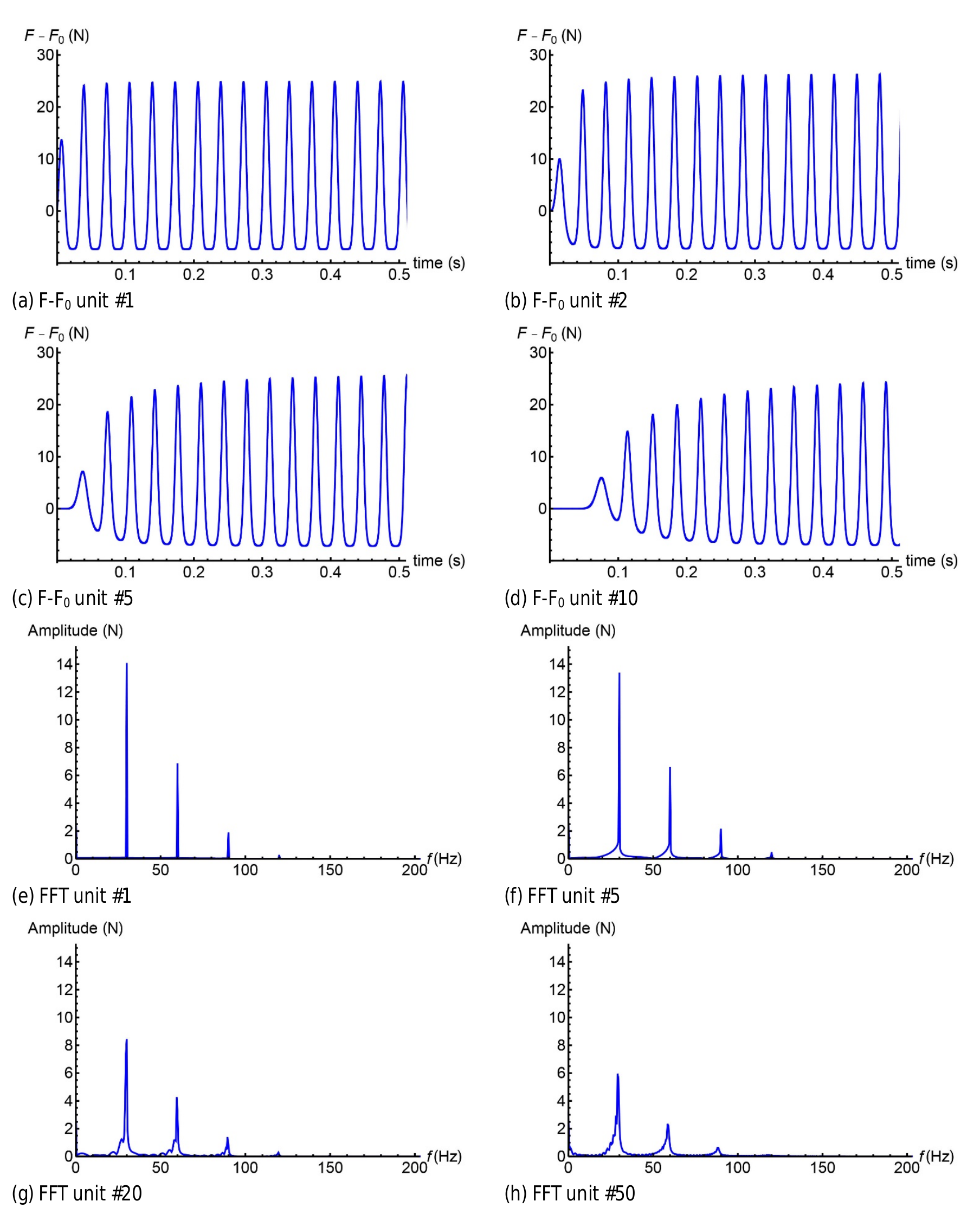}
	\caption{$\Delta F$ outputs in units \# 1 (a), \#2 (b), \# 5 (c), and \# 10 (d) of system Mono 1, and FFTs of outputs in units \# 1 (e), \# 5 (f), \# 20 (g), and \# 50 (h), which are induced by a sinusoidal time-displacement input with 0.03 mm amplitude and 30 Hz frequency.}
	\label{fig:panel30}
\end{figure}

\begin{figure}[htbp] 
	\centering
	\includegraphics[width=440pt]{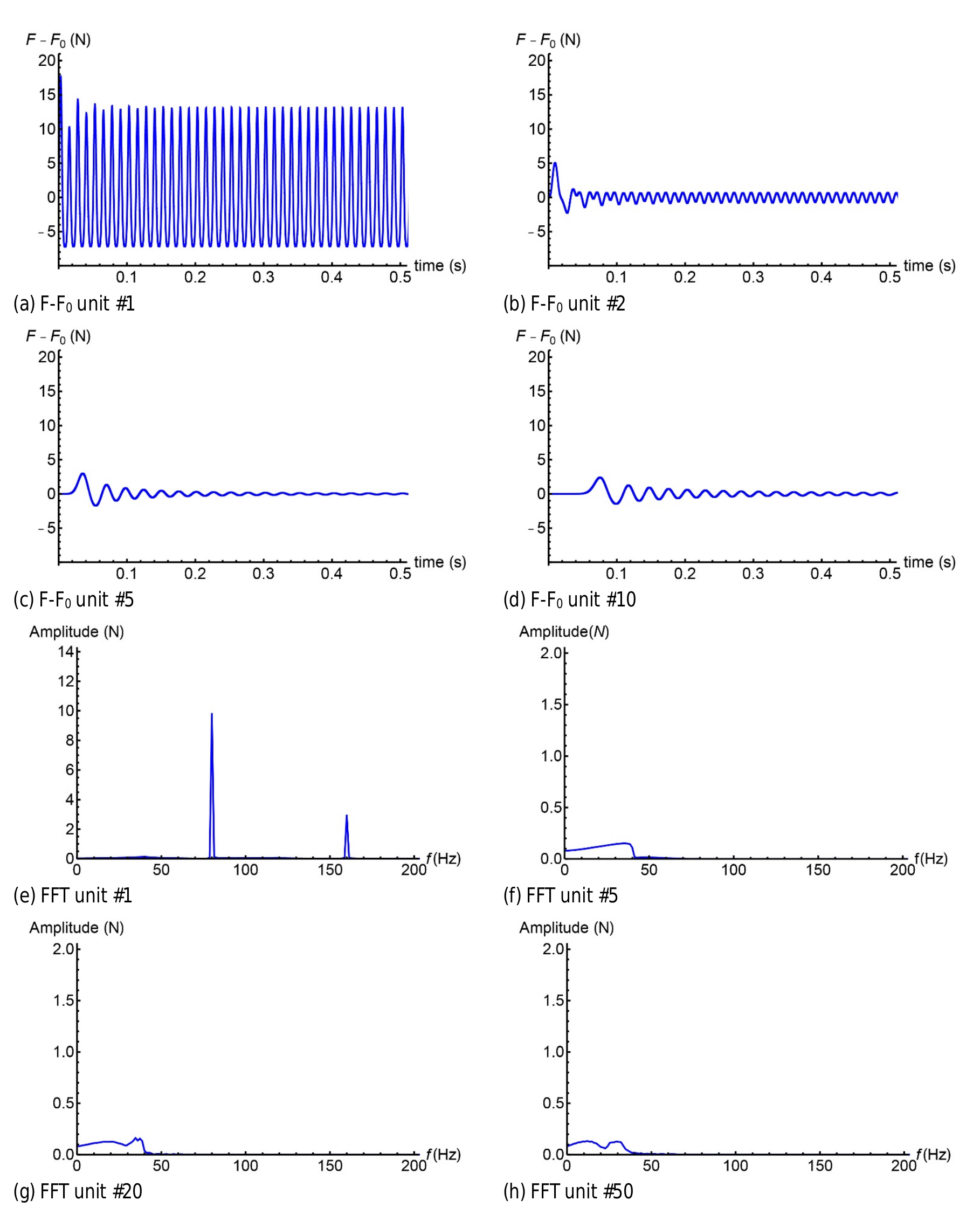}
	\caption{$\Delta F$ outputs in units \# 1 (a), \#2 (b), \# 5 (c), and \# 10 (d) of system Mono 1, and FFTs of outputs in units \# 1 (e), \# 5 (f), \# 20 (g), and \# 50 (h), which are induced by a sinusoidal time-displacement input with 0.03 mm amplitude and 80 Hz frequency.}	
	\label{fig:panel80}
\end{figure}

\begin{figure}[htbp] 
	\centering
	\includegraphics[width=450pt]{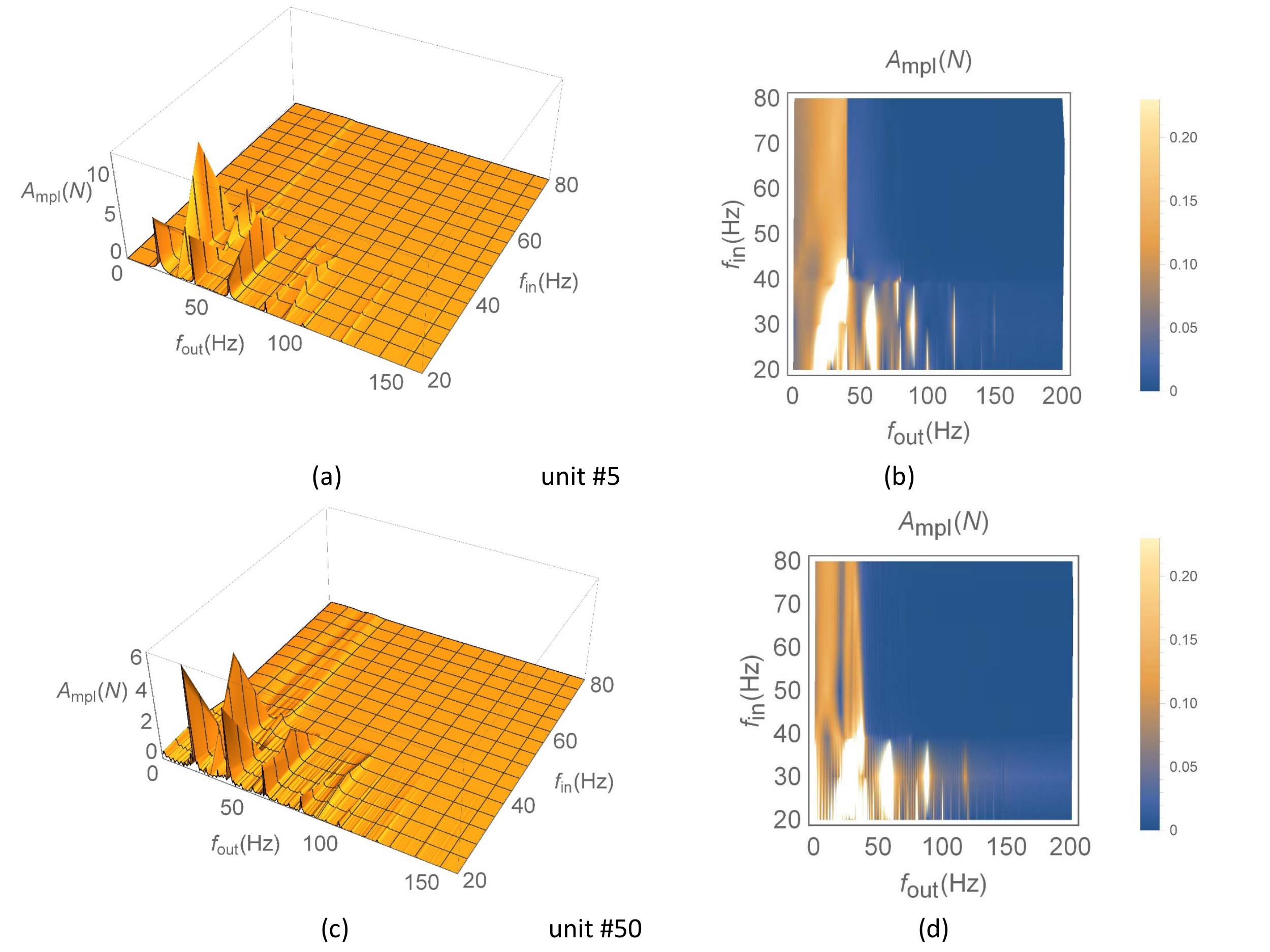}
	\caption{3D and density plots of the FFTs  of the outputs recorded in units \# 5 and \# 50 of system Mono 1  under sinusoidal time-displacement inputs with 0.03 mm amplitude and excitation frequency ranging between 20 Hz and 80 Hz.}
	\label{fig:FFTpanel}
\end{figure}

Figs. \ref{fig:panel30}-\ref{fig:panel80} illustrate the force vs. time outputs for the units 1, 2, 5 and 10 and the fast Fourier transforms (FFTs) of the  outputs for units 1, 5, 20 and 50 at excitation frequencies 30 Hz and 80 Hz, respectively. The FFT results are obtained through the Matlab\textsuperscript{\textregistered} function `fft' (Version R2017b) .
 The nonlinear response of the analyzed system is clearly visible, since one observes that the output force-time histories  $\Delta F = F - F_0$ feature positive peaks larger than the negative peaks, as a consequence of the stiffening-type response of the generic unit (cf. Fig. \ref{fig:Feps} a). The applied excitation induces transient oscillatory pulses $\Delta F$ followed by a steady state signal propagating throughout the chain. The latter is characterized by a leading harmonic with frequency ${f_{in}}$, and higher-order harmonics of ${f_{in}}$, and with reduced amplitude (cf. the (e) panels in  Figs. \ref{fig:panel30}-\ref{fig:panel80})
 \cite{SS2016}.
Fig. \ref{fig:panel30} shows that the input excitation of frequency 30 Hz {\cred{propagates unperturbed through the system}} (cf. panels (e)-(h)). 
Differently, Fig. \ref{fig:panel80}  shows that the input disturbance of frequency of 80 Hz generates a dramatically attenuated output. 
The $\Delta F$ output for ${f_{in}}=80$ Hz is indeed very fast reduced in amplitude as it travels along the chain, and progressively vanishes with time already at unit \# 2 (cf. panels (a)-(d) of Fig. \ref{fig:panel80}).
We observe that the FFT of the $\Delta F$ output at unit \# 5 exhibits almost zero amplitude for both ${f_{in}}$ and higher-order harmonics. The FFT plots for ${f_{in}}=80$ Hz in correspondence to the units 5, 20 and 50  feature nearly flat response, with amplitude approximatively equal to 0.1 N, in correspondence to the frequency range below the lower band gap edge of the linear Mono 1 chain ($40.03$ Hz). A similar, small amplitude plateau is present also in the FFT of the  $\Delta F$ output at unit \# 1 (not visible in Fig. \ref{fig:panel80}(e) because of its reduced amplitude), and is generated by the transient noisy response of the system.
%, already after after only a few tenths of seconds from the application of the disturbance.
The final Fig. \ref{fig:FFTpanel} illustrates 3D and density plots of the FFTs of the $\Delta F$ outputs recorded at units \# 5 and \# 50 of the nonlinear Mono 1 chain, as the excitation frequency varies between 20 Hz and 80 Hz. The results in  Fig. \ref{fig:FFTpanel} show that inputs with excitation frequency up $\sim 40$ Hz are allowed to propagate through the current system. Differently, inputs with excitation frequency greater than $40$ Hz are evanescent and generate outputs falling in dark/blue band gap regions of the density plots in Fig. \ref{fig:FFTpanel}. For excitation frequancies greater the 40 Hz, the plots in Fig. \ref{fig:FFTpanel} highlight only low FFT components outside the band gap region.
%{\crevv{  DA ELIMINARE: whose magnitude is about equal to 1/50-1/100  of the maximum FFT amplitude at unit \# 1 (0.1-0.2 N vs. 10-15 N, cf. panels (b,d) of Fig. \ref{fig:FFTpanel} and panels (e) of Figs. \ref{fig:panel30}-\ref{fig:panel80}).Despite}}
{\cred{T}}he presence of band gaps in the frequency spectrum is a property of linear systems (cf., e.g., Ref. \cite{Herbold}).{\cred{ However, in our system}} we observe that the presence of moderately large incremental strains does not substantially alter the structure of the dispersion curve shown in Fig. \ref{fig:monodispersion}.
%{\cred{.}} {\crevv{ DA ELIMINARE: which was obtained on assuming the Mono 1 chain excited by infinitesimally small perturbations of the equilibrium configuration \cite {ashmer}}}.

%%%%%%

%%%%%%

%%%%%%%%%%%%%%%%%%%%%%%%[SECTION: Concluding remarks]%%%
\section{Concluding remarks}\label{concl}

We have analyzed the frequency band structure of 1D tensegrity metamaterials formed by alternating tensegrity prisms with lumped masses. The conducted study assumed constant material properties (eventually accounting for units equipped with rigid bases and bars \cite{JPS20}), and variable states of local and global prestress of the system.
The results presented in Sect. \ref{th1} have shown that the examined structures exhibit highly tunable frequency band gaps in the linear regime induced by small vibration near the initial equilibrium state, as a function of a parameter $\bar p$ describing prestretching the cables in the tensegrity unit, and the initial strain $\eps_0$ induced by the precompression applied to the whole system.
By playing with such parameters it is shown to be possible to design monoatomic and biatomic systems that feature band gaps either in the audible and/or in the ultrasonic frequency range (cf. Sect. \ref{th1b}). In Sect. \ref{nonlin} we have generalized these results to a nonlinear regime induced by moderately large incremental strains for a monoatomic system with stiffening-type elastic response.
Both the analytic and numerical results presented in Sects. \ref{th1} and \ref{nonlin} have revealed a novel feature of tensegrity metamaterials, not previously investigated in the to-date literature (see  \cite{Skelton_2010}-\cite{Rimoli2017} and references therein), which consists of their ability to serve as band gap systems with easily tunable performance by playing with local and global prestress variables.

The present study paves the way to a number of relevant extensions and generalizations that we address to future work.
One natural extension of the current research regards the band structure of lattice metamaterials equipped with multi-atomic bases \cite{Theocharis,ashmer}, which can be richly designed by alternating tensegrity units equipped with different, material and prestress properties and lumped masses. Such metamaterials may function as stop band systems \cite{Theocharis,Herbold}, wave guides  \cite{Ruzzene:2005,Casadei:2013}, impact protection gear \cite{APL2014,21}, and/or acoustic lenses \cite{Spadoni2010,Donahue2014}.
Another relevant generalization of the present study regards the modeling of the dispersion behavior of tensegrity metamaterials in the nonlinear regime induced by large strains, to be conducted by recourse to particle dynamics simulations \cite{Herbold}, and/or the transfer matrix method (refer, e.g., to \cite{Hussein2014} and references therein).
Also the modeling of the mechanical response and band structure of 3D tensegrity metamaterials deserves special attention, which requires the use of numerical codes dedicated to the dynamics of spatial tensegrity systems \cite{3dtensegrity}, and/or finite elements simulations \cite{pentamodeband}.
Finally, the additive manufacturing and the experimental testing of physical models of tensegrity metamaterials at different scales is a topic of great interest and a challenge at present \cite{Amendola_2015}, since it requires the employment of advanced multimaterial deposition techniques that can handle internal prestress. One viable strategy consists of using projection micro-stereolitography setups \cite{Howon1} that employ swelling materials for the tensile members \cite{Howon2}. Once dried, these members will contract, creating internal prestress.
Alternatively, one can use multi-jet  technologies that handle  materials with different coefficients of thermal expansion for struts and cables, in order to create internal self-stress during the deposition process.

\section*{Acknowledgements}
A.A. gratefully acknowledges financial support from the Department of Civil Engineering at the University of Salerno.
A.O.K. thanks the funding from the European Union's 7th
Framework programme for research and innovation under the Marie Sk{\'l}odowska-Curie Grant Agreement No. 609402-2020 researchers: Train to Move (T2M). 
%C.D. thanks (insert here...). N.M.P. is supported by the European Research Council (ERC PoC 2015 SILKENE No. 693670), and by the European Commission under the Graphene FET Flagship (WP14 `Polymer Nanocomposites' No. 604391) and FET Proactive `Neurofibres' grant No. 732344. 
%F.F. thanks (insert here...). 
N.M.P. is supported by the European Commission H2020 under the Graphene Flagship Core 1 No. 696656 (WP14 `Polymer Composites') and FET Proactive `Neurofibres' grant No. 732344.

\section*{Compliance with Ethical Standards}
The authors declare that they have no conflict of interest.

\end{document}